\documentclass[a4paper,fleqn,usenatbib]{mnras}

\usepackage{newtxtext,newtxmath}

\usepackage[T1]{fontenc}
\usepackage{ae,aecompl}

\usepackage{graphicx}	
\usepackage{amsmath}	
\usepackage{amssymb}	

\title[Neural network-based analysis of X-ray spectra]{Neural network-based preprocessing to estimate the parameters of the X-ray emission of a single-temperature thermal plasma}

\author[Y. Ichinohe et al.]{
Y. Ichinohe,$^{1}$\thanks{E-mail: ichinohe@tmu.ac.jp}
S. Yamada,$^{1}$
N. Miyazaki,$^{1}$
and S. Saito$^{2}$
\\
$^{1}$Department of Physics, Tokyo Metropolitan University, Minami-Osawa 1-1, Hachioji, Tokyo\\
$^{2}$Department of Physics, Rikkyo University, Nishi Ikebukuro 3-3-4-1, Toshimaku, Tokyo\\
}

\date{Accepted 2018 January 17. Received 2018 January 16; in original form 2017 November 02}

\pubyear{2018}

\begin{document}
\label{firstpage}
\pagerange{\pageref{firstpage}--\pageref{lastpage}}
\maketitle

\begin{abstract}
We present data preprocessing based on an artificial neural network to estimate the parameters of the X-ray emission spectra of a single-temperature thermal plasma. The method finds appropriate parameters close to the global optimum. The neural network is designed to learn the parameters of the thermal plasma (temperature, abundance, normalisation, and redshift) of the input spectra. After training using 9000 simulated X-ray spectra, the network has grown to predict all the unknown parameters with uncertainties of about a few percent. The performance dependence on the network structure has been studied. We applied the neural network to an actual high-resolution spectrum obtained with {\it Hitomi}. The predicted plasma parameters agreed with the known best-fit parameters of the Perseus cluster within $\lesssim10$\% uncertainties. The result shows a possibility that neural networks trained by simulated data can be useful to extract a feature built in the data, which would reduce human-intensive preprocessing costs before detailed spectral analysis, and help us make the best use of large quantities of spectral data coming in the next decades.
\end{abstract}




\section{Introduction}

In X-ray astronomy, model-fitting is widely used to extract physical implications such as model parameters from given spectral data. Finding the global optimum of an objective function, such as the minimum of $\chi^2$ or the maximum of a likelihood function, is one of the major difficulties in model-fitting. Typical optimization algorithms such as the Nelder--Mead method \citep{nelder65} or the Levenberg--Marquardt algorithm \citep{more78} are designed to find only a local optimum, and thus complex spectral features of the model such as narrow emission lines make the fitting more prone to fall into a local minimum compared to featureless continuum models. In such a case, choosing appropriate initial parameters based on enough knowledge about the model and experience of spectral fitting is required to help these algorithms to find the global optimum.

Even if one acquires such a skill, the number of spectra will increase by a few orders of magnitude over the next decades -- {\it Athena} will have $\sim$4000 pixels of TES (Transition Edge Sensor) calorimeters \citep{nandra13}, and further future calorimeter missions such as {\it Lynx} \citep{gaskin16} and {\it DIOS} \citep{yamada16} propose a few tens of thousands of pixels. Therefore, data preprocessing will be an essential step in mining large amounts of data in future, and thus developing an automatic way to find appropriate initial parameters is necessary to accommodate a growing number of pixels of the microcalorimeters.

A frequently-used way to find the global minimum in a multi-dimensional parameter space of the objective function is the method of grid search. Standard routines in X-ray astronomy, such as {\small XSPEC} \citep{arnaud96}, are usually equipped with such a subroutine, which provides a function of moving parameters to understand the shape of the multi-dimensional surface in the parameter space (e.g. {\tt steppar} command in {\small XSPEC}). Although such an empirical method has worked for a handful of data, it will face the limit of the computational time over the next few decades. Let the number of parameters, grids, and model components, be $n$, $m$, and $c$, respectively, the amount of calculation would scale as $\sim (m^n)^c$. The exponential scaling would inevitably lead to an increase of computational cost.

Exploiting the prior knowledge obtained in different ways, e.g. using the redshift obtained from optical spectroscopy as the plasma redshift of a galaxy cluster, or using the best fit of a neighbouring pixel, probably helps well-behaved initial guesses of the parameters to be set. However, such methods might not work if for example line-of-sight velocity shear exists in the galaxy cluster, which is often expected due to cold fronts, stripped tails, or filaments. Therefore, a complementary method which can predict the initial parameters independently from other knowledge would be important.

An artificial neural network, which predicts initial parameters that are sufficiently close to the most optimal parameters from the dataset itself, might resolve the problem. Artificial neural network is one of machine learning techniques, which consists of multiple formal neurons (computational model of real human neurons) to mimic the operations of human brain to achieve good performance in cognitive tasks \citep[see e.g.][for elaborate explanations]{lecun15,goodfellow16}. The idea dates back to 1940s, but the high computational cost to train such artificial neural networks has been the main problem that prevents them from being popular. The recent evolutionary computational progress, especially the growth of Graphics Processing Units (GPUs) has made it possible to actually train the neural network in a reasonable time, making the technique very popular in many areas. These areas include e.g. image recognition \citep[][]{krizhevsky12}, natural language processing \citep[][]{mikolov13} and games \citep[][]{silver16} as practical applications, and also astrophysical image analysis \citep[][]{hezaveh17,kimura17,caron17}, light curve analysis \citep[][]{shallue17} and condensed matter physics \citep[][]{tanaka17} as scientific applications.

Once trained, such an artificial neural network would automatically give the set of the parameters almost instantly. This reduces the initial trial-and-error process, which is significant in the big-data era of the next decades. Most of the machine learning techniques necessitate a training and evaluation dataset of significant size (supervised learning). Fortunately, the X-ray spectra in most cases can be infinitely generated by using the detector response matrices and a model that can be calculated by specifying required parameters.

A similar concept has been considered by \citet{larsen92}, but was not practical at that time. In this paper, we design the neural network to learn the parameters of single-temperature thermal plasma (temperature, abundance, normalisation, and redshift) of the input spectral data, and actually construct and train it to introduce neural networks as a potential option for data processing automation in the near future, by demonstrating that the network works on the real astrophysical data. The setup of the network is shown in Section~\ref{sec:nn}. By training the network several hundred times, it has grown to predict all the unknown parameters with uncertainties of about a few percent. The dependence of the network performance on its depth and width has been investigated. The performance for actual observational data is confirmed by using the high energy-resolution spectrum taken with the {\it Hitomi} satellite \citep{takahashi16,hitomi16}, resulting in a good agreement with the best-fitting values. These results are shown in Section~\ref{sec:result}. The applicabilities and prospects of machine-learning preprocessing before a canonical spectral analysis are discussed in Section~\ref{sec:discussion}.

\section{The neural network}\label{sec:nn}
\subsection{Plasma model and training dataset}

A training dataset is prepared for the supervised learning of the neural network. We used the {\tt fakeit} command in {\small XSPEC} to simulate the instance of each given combination of plasma parameters. As the input of the {\tt fakeit} command, we used the response files used in the data analysis of the {\it Hitomi} Perseus observation \citep[e.g.,][]{ichinohe17,nakashima17,noda17}. The spectra are simulated with the exposure time of 1~Msec.

In the simulation, we adopted the model of a single-temperature thermal plasma in collisional ionization equilibrium, attenuated by the Galactic absorption; {\tt TBabs*bapec} \citep{smith01}. 10000 spectra are randomly generated with two fixed parameters and four randomly generated parameters as shown in the list below:
\begin{itemize}
\item the neutral hydrogen column density $N_H$: $1.38\times10^{21}$~cm$^2$ (fixed to the value obtained in Leiden/Argentine/Bonn (LAB) survey \citep{kalberla05})
\item temperature $kT$ : 1.0--10.0~keV
\item Fe abundance $Z$ : 0.1--1.5~solar
\item redshift $z$ : 0.0--0.1
\item velocity broadening $v$ : 160.0~km~s$^{-1}$ (fixed)
\item normalisation $N$ : 0.01--1.0,
\end{itemize}
where $kT$, $Z$ and $z$ are randomly generated from a uniform distribution in the above ranges, while $N$ is sampled in a log-uniform distribution. Note that, from the spectral point of view, $kT$ changes both the continuum shape of bremsstrahlung and the ionization balance of heavy elements which leads to changes in emission line distributions, while $Z$ changes the strengths of the lines. Changes in $z$ appear as a linear transformation ($E \rightarrow E/(1+z)$) along the energy axis.

\subsection{Network design}

\begin{figure}
\begin{center}
\includegraphics[width=0.48\textwidth]{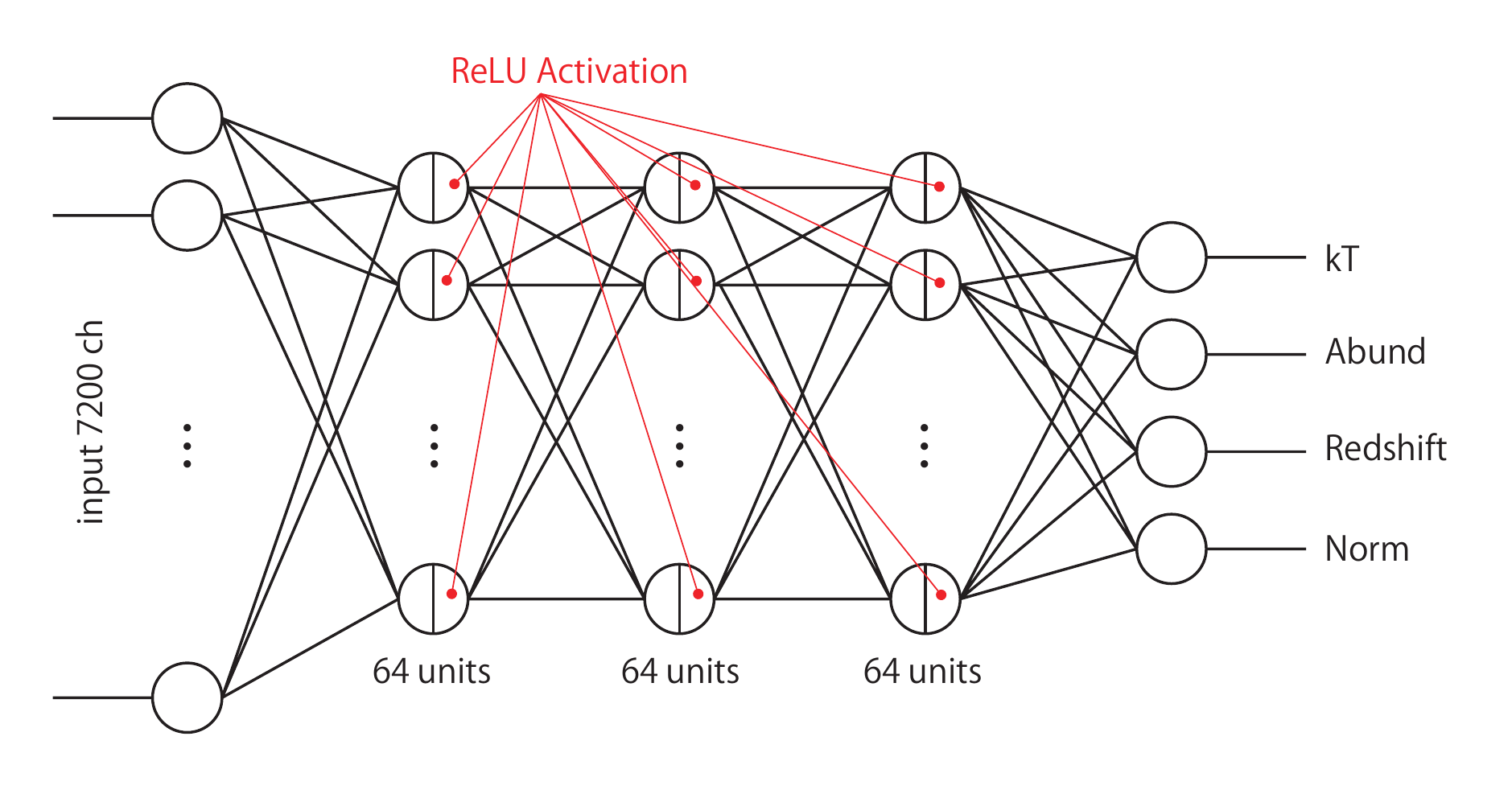}
\caption{Diagram of the neural network. The network structure of the example is 7200-64-64-64-4. Each circle represents a ``neuron'' in the network. See the main text for the details of the network.}
\label{fig:network}
\end{center}
\end{figure}

Our concept of the neural network is shown in Figure~\ref{fig:network}. Each circle represents a ``neuron'' in the network. A neuron has $n$ inputs $x_i$ and one output $y$. The $n$ inputs are linearly combined by tunable parameters and nonlinearly transformed; $y=f(\sum^{n}_{i} w_ix_i + b)$, where $i$ is the index of the input, $w_i$ is the $i$th weight parameter, $b$ is the bias parameter, and $f(x)$ is a nonlinear function of $x$ (activation function). Such neurons are connected in a layered manner (a vertical series of the neurons in Figure~\ref{fig:network}), where all the neurons in a layer shares the same input, and all the outputs from the neurons in this layer are used as the input to the next layer (fully connected). The structure of the network is called as a multilayer perceptron.

The input to the first layer is a 1.8--9.0~keV X-ray count spectrum with the energy bin size of 1~eV, i.e., a single 7200-dimensional vector. The maximum count among all the energy bins of all the simulated spectra was 60963, and thus the input vector is scaled by $1/10^5$ to restrict all the input values between 0 and 1. The output of the first layer is connected to the next, fully connected layer. One to four hidden layers, which are all fully connected layers, are present before the last layer. Each hidden layer neuron is activated using a Rectified Linear Unit \citep[ReLU;][]{nair10}, which is a very commonly used activation function ($f(x)$). The output of the last layer is a four-dimensional vector, each component of which corresponds to each one of the plasma parameters shown in the previous section. These output variables are converted so that every parameter is uniformly distributed in the range of 0--1 in order to avoid a biased learning toward the parameters with higher impacts. The bias originates from the fact that variables with higher numerical values have higher impacts on the loss function if the outputs are not normalised.\footnote{For example, the difference of the numerical values of temperature between 9~keV and 10~keV is 1, which is ten times the difference of the numerical values of Fe abundance between 0.9~solar and 1.0~solar of 0.1. This means that, without normalisation, and with the loss function of mean squared error in our case, a 10\% change of temperature has one hundred times larger impact on the loss function compared to a same 10\% change of Fe abundance.}

9000 out of 10000 simulated spectra are used for the network training, while the rest are used for the evaluation of the network performance. We used the high-level neural network library Keras \citep{chollet15} with the TensorFlow backend \citep{abadi15}. The loss function was set to be the mean squared error between the output and the training data. Adaptive Moment Estimation \citep[Adam;][]{kingma14} was used for optimization. The batch size was 50, and the training was repeated only for 1000 epochs because the value of the loss function converged before $\sim$500 epochs.\footnote{Sample codes are available at \url{https://github.com/yutoichinohe/sample_codes}.}

\begin{table*}
 \caption{Summary of the neural networks and their results. }
 \label{tab:network}
   \begin{center}
   \begin{tabular}{lrrrrrc} \hline \hline
Network$^a$      & best loss$^b$  & $\Delta T_{68}$(\%)$^c$ & $\Delta Z_{68}$(\%)$^c$ & $\Delta z_{68}$(\%)$^c$ & $\Delta N_{68}$(\%)$^c$ & Remark      \\ \hline  
%
128-128-128-128-4 &  1.1$\times10^{-4}$ & 1.2  & 2.0  & 1.8  & 1.7  & \\
128-128-128-4     &  1.3$\times10^{-4}$ & 1.4  & 2.2  & 1.9  & 2.2  & \\
128-128-4         &  2.2$\times10^{-4}$ & 1.7  & 3.2  & 2.8  & 2.4  & \\
128-4             &  6.0$\times10^{-4}$ & 2.9  & 5.2  & 4.4  & 4.2  & \\[0.2em]
64-64-64-64-4     &  1.6$\times10^{-4}$ & 1.6  & 2.6  & 2.7  & 2.4  &\\
64-64-64-4        &  2.5$\times10^{-4}$ & 1.8  & 3.0  & 2.7  & 2.8  & used in figure  \\
64-64-4           &  4.2$\times10^{-4}$ & 2.4  & 4.3  & 4.6  & 3.3  & \\
64-4              & 10.9$\times10^{-4}$ & 5.0  & 8.1  & 7.3  & 8.5  & \\[0.2em]
32-32-32-32-4     &  3.6$\times10^{-4}$ & 2.5  & 3.7  & 4.0  & 3.5  &  \\
32-32-32-4        &  4.4$\times10^{-4}$ & 2.3  & 4.7  & 4.4  & 3.8  & \\
32-32-4           &  8.1$\times10^{-4}$ & 3.7  & 6.3  & 6.5  & 6.3  & \\
32-4              & 19.3$\times10^{-4}$ & 6.1  & 10.2 & 8.5  & 9.4  & \\[0.2em]
16-16-16-16-4     &  9.3$\times10^{-4}$ & 3.4  & 6.6  & 8.0  & 5.3  &  \\
16-16-16-4        & 14.2$\times10^{-4}$ & 6.3  & 8.7  & 8.2  & 9.0  & \\
16-16-4           & 20.6$\times10^{-4}$ & 7.5  & 11.0 & 10.6 & 10.2 & \\
16-4              & 58.7$\times10^{-4}$ & 10.4 & 19.0 & 19.7 & 17.9 & \\

\hline \hline 
   \end{tabular}
   \end{center}
\begin{itemize}
\item[$^a$] The dimension of the input vector is 7,200 in all cases.
\item[$^b$] The best attained value of the loss function among sixteen trials.
\item[$^c$] For the definition of $\Delta x_{68}$, see the main text.
\end{itemize}
\end{table*}

We have tried several configurations of the network by changing the depth (the number of the hidden layers) and width (the number of the hidden units in a layer). The patterns are shown in Table~\ref{tab:network}. The maximum width of the first layer is set at 128. This width limit is empirically derived by performing a principal component analysis for the first 7200$\times$256 weight matrix in a 7200-256-64-16-4 network, which accepts a single 7200-dimensional vector as an input, has three fully-connected hidden layers with 256, 64, and 16 hidden units, and returns a four-dimensional vector as an output. The contributing rate steeply decreases from $\sim$0.5 to $\sim$10$^{-4}$, while the decline becomes almost flat below $\sim$10$^{-4}$. The number of components that have a contribution rate above $\sim$10$^{-4}$ is about 100. Note that even when the different datasets, such as those with different plasma parameter ranges, are used to create the input data, the number of components was close to 100.

\section{Result}\label{sec:result}

\subsection{Parameter estimation}

\begin{figure*}
\begin{center}
\includegraphics[width=0.98\textwidth]{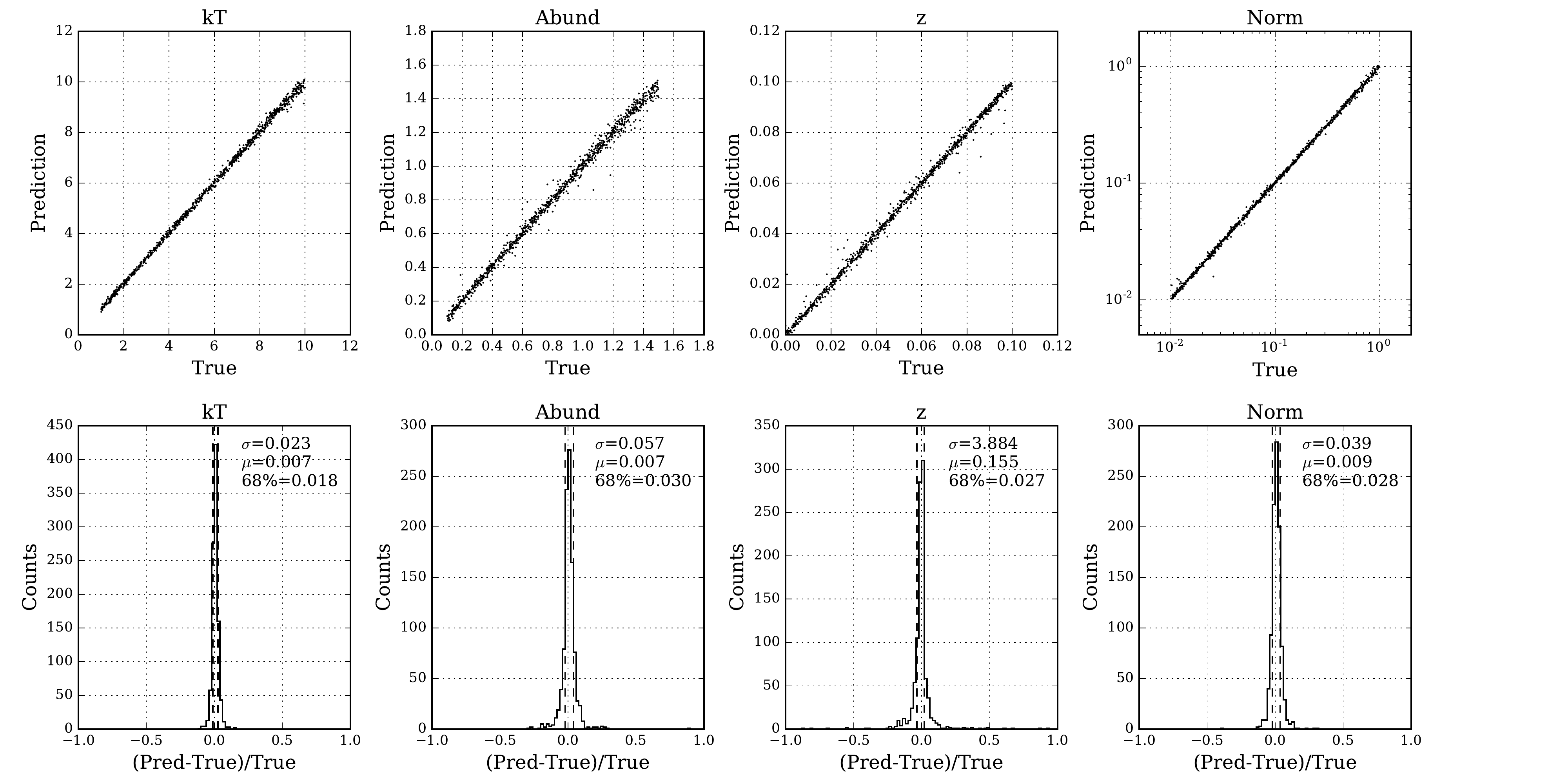}
\caption{{\it Top:} correlation between the input parameters and the neural network predictions. {\it Bottom:} the relative deviations of the predicted values from the input parameter.}
\label{fig:paraplot}
\end{center}
\end{figure*}

We here describe the result obtained by the 7200-64-64-64-4 network. Figure \ref{fig:paraplot} shows the result of the parameter estimation using the network that attained the minimum value of the loss function among sixteen trials with the same network and shuffled dataset. We evaluated the reproducibility of the parameters by $\Delta x_{68} \equiv (\Delta x_{84}-\Delta x_{16})/2$, where $\Delta x_p$ is the $p$\% percentile of $\Delta x$, which is defined by the formula: $\Delta x \equiv (x_{\rm pred}-x_{\rm true})/x_{\rm true}$, with $x_{\rm pred}$ and $x_{\rm true}$ being the predicted and the input values of the variable $x$, respectively. Although the sigma of the best-fit Gaussian or the standard deviation are alternative choices to $\Delta x_{68}$, we chose this because it can be calculated purely from the data and is robust against the outliers of the predicted values. The reproducibility is $\lesssim$3\% for all the four output parameters. Although the range of the parameter values and how the model depends on the parameter are different for each parameter, the attained reproducibilities are similar within a factor of 2. This is probably because the renormalisation and conversion of the input and output parameters reduce the diversities.

We also studied the dependence of the network performance on the statistics of input data. We simulated X-ray spectra in the same way as making the training datasets, with lower exposure times of 100 ($1/10$) and 10~ksec ($1/100$). 1000 spectra were realised for each exposure time. The obtained $\Delta T_{68}, \Delta Z_{68}, \Delta z_{68}$ and $\Delta N_{68}$ are 2.8\%, 7.2\%, 4.3\% and 3.6\% for the 100~ksec dataset, and 6.0\%, 14.2\%, 10.2\%, and 7.1\% for the 10~ksec dataset. We thus conclude that the network works also on the data with lower statistics, with natural degradation regarding the input data quality.

When the focus is moved onto the outlier events, $Z$ and $z$ are slightly more distributed at the rim of the distribution. Such unsuccessful events tend to have lower $kT$ so that many lines do not show up distinctively. The reason for the failure in finding the true $Z$ and $z$ is probably because the spectra are generated with the {\it Hitomi} response and auxiliary files that has a strong absorption due to the gate valve at low energies. However, such outlier events that deviate from the true values by 20\% are quite few, below 5\% of the total events. Thus, in contrast to visual inspection or empirical trials, this network that gives us an initial guess of the parameters at the accuracy of a few percent can accelerate to grasp the features of lots of spectra.

\subsection{Performance and the network design}

We further studied the dependence of the network performance on hyperparameters such as network depth or width to see if there are any better results obtained or how they differ by the choice (Table~\ref{tab:network}). The tried widths are 128, 64, 32, and 16. The number of hidden layers is 1 to 4. The number of the hidden units is the same in all the hidden layers. For the same number of hidden layers, a wider network works better. Comparing constant width, deeper is better. In short, a bigger network worked better, although the improvement of the performance due to adding another extra layer is relatively less significant for deeper networks. This indicates that the performance improvement is saturated by the limit determined from the training dataset.

To estimate the reproducibility of the network performance after the same set of training, we shuffled the entire dataset sixteen times. Each time the dataset is split in 9000 and 1000, and same training procedure was performed, using the 9000 for training and the 1000 for evaluation. We find that the best attained value of the loss function can be different by a factor of $\sim$5, with the typical difference being a factor of $\sim$2.

Using a commercial graphics card, NVIDIA GeForce GTX 1080 Graphics Card, the typical time, GPU memory and the number of CUDA (Compute Unified Device Architecture) cores needed to train a single network were 7--12~minutes, 200--300~MiB, and less than 1000, respectively, depending on the scale of the network.

\subsection{Application to the observed data}

\begin{figure*}
\begin{center}
\includegraphics[width=0.98\textwidth]{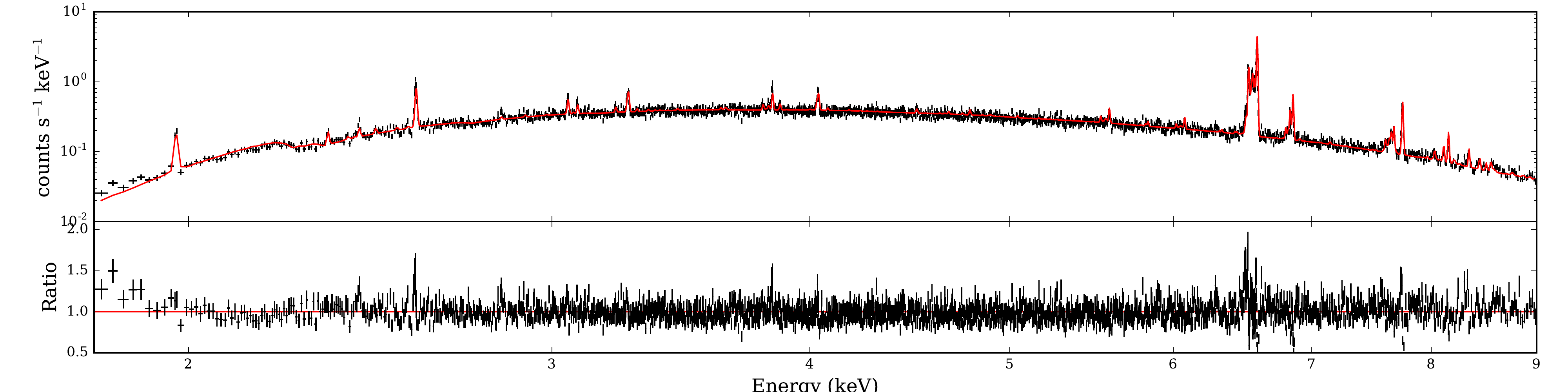}
\includegraphics[width=0.98\textwidth]{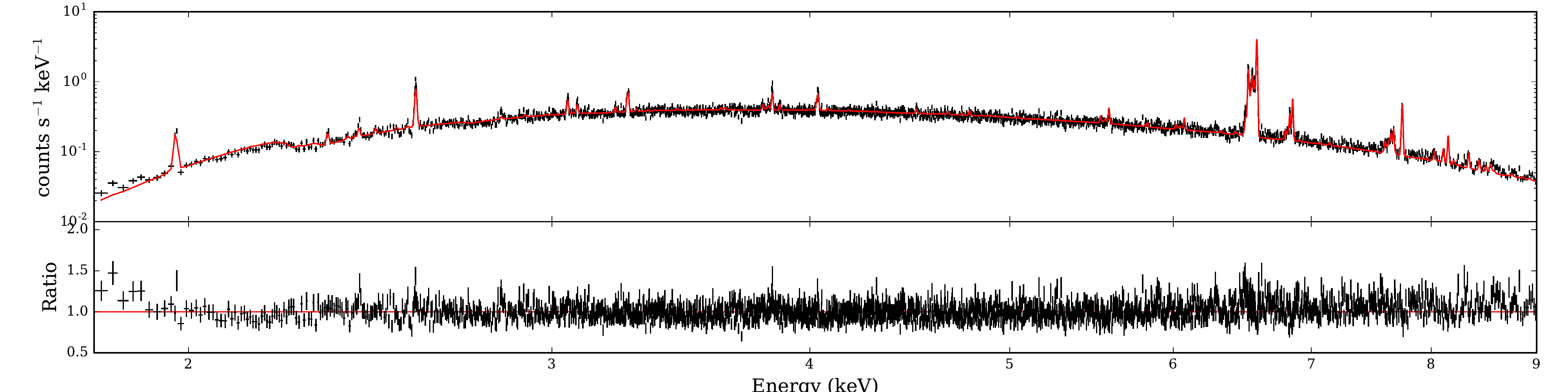}
 \caption{{\it Top:} the model calculated based on the values predicted by the neural network for the actual {\it Hitomi} spectrum. {\it Bottom:} the best-fit model found using the initial parameter estimation to be the predicted values above. Red curves are the models, and black data points are the {\it Hitomi} spectrum and residuals. Note that the actual obtained spectrum (in the form of counts in each energy bin) is normalised by the exposure time only for display.}
\label{fig:hitomi}
\end{center}
\end{figure*}

We finally apply our network to actual high-resolution spectroscopic data. Figure~\ref{fig:hitomi} shows the {\it Hitomi} microcalorimeter \citep[SXS; Soft X-ray Spectrometer,][]{kelley16} spectrum of the core of the Perseus cluster taken during {\it Hitomi}'s initial commissioning phase. Only the data of Obs~3 (ObsIDs 10040030, 10040040 and 10040050) are used and the parabolic gain correction is applied to the data \citep[see Appendix~1 of][for details]{nakashima17}. The net exposure is 146~ksec (0.146~Msec). The input {\it Hitomi} spectrum was therefore first scaled by $1/0.146$ to match the simulation input, then scaled by $1/10^5$.

The predictions of the network were $kT = 4.39$~keV, $Z = 0.574$~solar, $z = 0.0168$, and $N = 0.184$. The spectrum computed using the predicted parameters and its residuals from the data are shown in the top panel of Figure~\ref{fig:hitomi}. Importantly, by performing the fitting with the initial parameters using the above values, the fit converges on the true global optimum as shown in the bottom panel of Figure~\ref{fig:hitomi}, demonstrating the effectiveness of the neural network for the initial preprocessing of the high-resolution spectroscopic data.

The best-fit parameters obtained for the same model as the simulation (i.e., {\tt TBabs*apec}) are $kT = 4.23\pm0.02$~keV, $Z = 0.518\pm0.005$~solar, $z = 0.01729\pm0.00001$, and $N = 0.190\pm0.001$. The predicted values are consistent with the best-fitting values within $\lesssim$10\%. The reason of the prediction being not as good as the performance shown in Table~\ref{tab:network} is likely because of the presence of another power-law continuum component and the neutral iron emission originating from the brightest cluster galaxy NGC~1275 \citep{noda17}. Despite this contamination, the network returned the parameters with enough accuracy to make the fitting find the global minimum. Therefore we conclude that with contamination of this level, the network works in practice.

\section{Discussion and summary}\label{sec:discussion}

We have constructed and tested a neural network to predict the parameters of a single-temperature thermal plasma. The actual spectra may be much more complicated and there should be many issues related to calibration, atomic data and so on. However even in those cases, the simplest modelling such as modelling with single-temperature plasma using knowledge of atomic physics at that moment would be an important first step, and even this step would suffer from the growing amount of future data. Also, even in the ideal (perfect calibration and purely single temperature plasma) case, the same big-data problem will occur. We think it is important to make this process more efficient.

 The network learned the spectral features that are sensitive to temperature, abundance, redshift, and normalisation, and as a result could estimate them within a few percent accuracy. This allows us to set the initial guess of the parameters to be the derived ones, which leads to a quick convergence on the global minimum in the multi-dimensional surface of the objective function, saving time for the initial trial-and-error processes in X-ray spectral analysis. Importantly, the network is reusable; the parameters are easily distributable in a binary format such as a CALDB file. Once such a pre-trained network (e.g. trained using typical response files) is distributed (e.g. officially from the calibration team), it would save a lot of individual human work.

Currently, choosing the model and providing initial values for a fit requires human interaction. Neural networks could also be used in data analysis pipelines to do this quick-look analysis in an automated way, providing a starting point with which human astrophysicists could do a more detailed and accurate analysis.

In general, each network has its specific task, determined by the policy with which the network is trained. In our case, we trained the network so that it is able to predict the parameters of one temperature plasma, and thus it cannot do anything else as it is. Training a network to do more complex tasks is probably possible, but it would be more difficult to properly train the network and would require more training datasets.

Using neural network in science has often been a matter of debate. The major issue is that neural networks are in general `black box' because it is difficult to intuitively understand the behaviour of the network compared to written-down algorithms. Unveiling what a `black box' does is a significant problem, while there have been attempts on it \citep[e.g.,][]{zeiler13}. As an attempt to understand how our network acquired the ability to predict the underlying parameters, we inspected the weight matrices of the hidden layers. We found that at the first layer, in some of the rows of the weight matrix (which has 7200 elements corresponding to 1.8--9~keV), the elements appear excited at the energies corresponding to emission lines. It seems that some important features especially related to the emission lines are learned in the early phase. As the stage of the network moves to the output, the matrix components seem featureless, which means either that the features learned in the deeper layers are complex or that the features are simply not apparent as they are shuffled because the role of each hidden unit is randomly assigned in the course of training.

In response to the recent development of the technique of deep convolutional neural network (CNN) and its application to image recognition \citep[see e.g.,][]{krizhevsky12}, there have already been many applications to a variety of astrophysical themes: e.g., strong gravitational lensing \citep[][]{hezaveh17}; supernovae \citep[][]{kimura17} or gamma-ray surveys \citep[][]{caron17}. Although spectral data are one-dimensional unlike image data, we think the technique of convolution is likely to be useful also for spectral data, because spectral lines are local structure and thus short-range correlations must exist.

In addition to the short-range correlation due to local structures such as emission lines, there should also be long-range correlations because for example some emission lines share the same origin; e.g, sometimes both the Fe-L complex and the Fe-K complex appear in the same spectrum while the former appears in the low-energy band and the latter appears in the high-energy band. In that sense spectral data are similar to natural languages rather than image data whose long-range correlation is not as strong as the short-range one, and therefore techniques for natural language processing such as recurrent neural network (RNN) or long short-term memory (LSTM) might be applicable.

Although we have demonstrated the effectiveness of the neural network for the regression of the parameters of a single-temperature plasma, the actual plasma may not always consist only of a single component. For example, the spectral data of the {\it DIOS} mission, which aims at the mapping of warm-hot intergalactic medium (WHIM), will consist of a few components of different plasma in different redshifts. We have tried to extend our methods to multi-component plasma -- classification by the number of plasma components. Adjusting hyperparameters, we so far achieved an accuracy of the classification of $\sim$80\%, which would not be sufficient for practical fitting routines. Such work that would require different techniques will be covered by the following paper.

Using the {\it Hitomi} data, \citet{ichinohe17} demonstrated the importance of associating velocity structures with morphological features. Future high-angular-resolution calorimeter missions such as {\it Lynx} will enable spatial mapping of velocity structures to be performed with a few orders of magnitude higher number of spatial bins. It is of vital importance to establish a data analysis strategy, which has a few orders of magnitude higher efficiency. Artificial neural network-based preprocessing could play an essential role to fully exploit the future, big astronomical data.

\section*{Acknowledgements}
We thank the referee for the constructive suggestions and comments. We thank Dr. Shinya Nakashima for the parabolic gain correction on the {\it Hitomi} data. YI is financially supported by a Grant-in-Aid for Japan Society for the Promotion of Science (JSPS) Fellows (16J02333). SY is supported by the JSPS Grant-in-Aid for Scientific Research 15H05438, 15H00785 and 16H03954, and NM is by 15K05024. SY and NM have worked on the micro calorimeter, and YI and SS have contributed to hard X-ray/gamma-ray detectors on board the {\it Hitomi} satellite. This work is supported by Astro-AI working group in RIKEN iTHEMS.





\begin{thebibliography}{}
\makeatletter
\relax
\def\mn@urlcharsother{\let\do\@makeother \do\$\do\&\do\#\do\^\do\_\do\%\do\~}
\def\mn@doi{\begingroup\mn@urlcharsother \@ifnextchar [ {\mn@doi@}
  {\mn@doi@[]}}
\def\mn@doi@[#1]#2{\def\@tempa{#1}\ifx\@tempa\@empty \href
  {http://dx.doi.org/#2} {doi:#2}\else \href {http://dx.doi.org/#2} {#1}\fi
  \endgroup}
\def\mn@eprint#1#2{\mn@eprint@#1:#2::\@nil}
\def\mn@eprint@arXiv#1{\href {http://arxiv.org/abs/#1} {{\tt arXiv:#1}}}
\def\mn@eprint@dblp#1{\href {http://dblp.uni-trier.de/rec/bibtex/#1.xml}
  {dblp:#1}}
\def\mn@eprint@#1:#2:#3:#4\@nil{\def\@tempa {#1}\def\@tempb {#2}\def\@tempc
  {#3}\ifx \@tempc \@empty \let \@tempc \@tempb \let \@tempb \@tempa \fi \ifx
  \@tempb \@empty \def\@tempb {arXiv}\fi \@ifundefined
  {mn@eprint@\@tempb}{\@tempb:\@tempc}{\expandafter \expandafter \csname
  mn@eprint@\@tempb\endcsname \expandafter{\@tempc}}}

\bibitem[\protect\citeauthoryear{Abadi et~al.,}{Abadi et~al.}{2015}]{abadi15}
Abadi M.,  et~al., 2015, {TensorFlow}: Large-Scale Machine Learning on
  Heterogeneous Systems, \url {https://www.tensorflow.org/}

\bibitem[\protect\citeauthoryear{{Arnaud}}{{Arnaud}}{1996}]{arnaud96}
{Arnaud} K.~A.,  1996, in {Jacoby} G.~H.,  {Barnes} J.,  eds,  Astronomical
  Society of the Pacific Conference Series Vol. 101, Astronomical Data Analysis
  Software and Systems V. p.~17

\bibitem[\protect\citeauthoryear{{Caron}, {G{\'o}mez-Vargas}, {Hendriks}  \&
  {Ruiz de Austri}}{{Caron} et~al.}{2017}]{caron17}
{Caron} S.,  {G{\'o}mez-Vargas} G.~A.,  {Hendriks} L.,   {Ruiz de Austri} R.,
  2017, preprint, \href {http://adsabs.harvard.edu/abs/2017arXiv170806706C} {}
  (\mn@eprint {arXiv} {1708.06706})

\bibitem[\protect\citeauthoryear{Chollet et~al.}{Chollet
  et~al.}{2015}]{chollet15}
Chollet F.,  et~al., 2015, Keras, \url{https://github.com/fchollet/keras}

\bibitem[\protect\citeauthoryear{{Gaskin}, {{\"O}zel}  \& {Vikhlinin}}{{Gaskin}
  et~al.}{2016}]{gaskin16}
{Gaskin} J.,  {{\"O}zel} F.,   {Vikhlinin} A.,  2016, in Space Telescopes and
  Instrumentation 2016: Optical, Infrared, and Millimeter Wave. p. 99040N,
  \mn@doi{10.1117/12.2240459}

\bibitem[\protect\citeauthoryear{Goodfellow, Bengio  \& Courville}{Goodfellow
  et~al.}{2016}]{goodfellow16}
Goodfellow I.,  Bengio Y.,   Courville A.,  2016, Deep Learning.
MIT Press

\bibitem[\protect\citeauthoryear{{Hezaveh}, {Levasseur}  \&
  {Marshall}}{{Hezaveh} et~al.}{2017}]{hezaveh17}
{Hezaveh} Y.~D.,  {Levasseur} L.~P.,   {Marshall} P.~J.,  2017, \mn@doi [\nat]
  {10.1038/nature23463}, \href {http://ads.nao.ac.jp/abs/2017Natur.548..555H}
  {548, 555}

\bibitem[\protect\citeauthoryear{{Hitomi Collaboration} et~al.,}{{Hitomi
  Collaboration} et~al.}{2016}]{hitomi16}
{Hitomi Collaboration} et~al., 2016, \mn@doi [\nat] {10.1038/nature18627},
  \href {http://ads.nao.ac.jp/abs/2016Natur.535..117H} {535, 117}

\bibitem[\protect\citeauthoryear{{Hitomi Collaboration} et~al.,}{{Hitomi
  Collaboration} et~al.}{2017a}]{ichinohe17}
{Hitomi Collaboration} et~al., 2017a, preprint (\mn@eprint {arXiv}
  {1711.00240})

\bibitem[\protect\citeauthoryear{{Hitomi Collaboration} et~al.,}{{Hitomi
  Collaboration} et~al.}{2017c}]{noda17}
{Hitomi Collaboration} et~al., 2017c, preprint, \href
  {http://ads.nao.ac.jp/abs/2017arXiv171106289H} {} (\mn@eprint {arXiv}
  {1711.06289})

\bibitem[\protect\citeauthoryear{{Hitomi Collaboration} et~al.,}{{Hitomi
  Collaboration} et~al.}{2017b}]{nakashima17}
{Hitomi Collaboration} et~al., 2017b, preprint, \href
  {http://ads.nao.ac.jp/abs/2017arXiv171206612H} {} (\mn@eprint {arXiv}
  {1712.06612})

\bibitem[\protect\citeauthoryear{{Kalberla}, {Burton}, {Hartmann}, {Arnal},
  {Bajaja}, {Morras}  \& {P{\"o}ppel}}{{Kalberla} et~al.}{2005}]{kalberla05}
{Kalberla} P.~M.~W.,  {Burton} W.~B.,  {Hartmann} D.,  {Arnal} E.~M.,  {Bajaja}
  E.,  {Morras} R.,   {P{\"o}ppel} W.~G.~L.,  2005, \mn@doi [\aap]
  {10.1051/0004-6361:20041864}, \href
  {http://ads.nao.ac.jp/abs/2005A%26A...440..775K} {440, 775}

\bibitem[\protect\citeauthoryear{{Kelley} et~al.,}{{Kelley}
  et~al.}{2016}]{kelley16}
{Kelley} R.~L.,  et~al., 2016, in Space Telescopes and Instrumentation 2016:
  Ultraviolet to Gamma Ray. p. 99050V, \mn@doi{10.1117/12.2232509}

\bibitem[\protect\citeauthoryear{Kimura, Takahashi, Tanaka, Yasuda, Ueda  \&
  Yoshida}{Kimura et~al.}{2017}]{kimura17}
Kimura A.,  Takahashi I.,  Tanaka M.,  Yasuda N.,  Ueda N.,   Yoshida N.,
  2017, in 2017 IEEE 37th International Conference on Distributed Computing
  Systems Workshops (ICDCSW). pp 354--359, \mn@doi{10.1109/ICDCSW.2017.47}

\bibitem[\protect\citeauthoryear{{Kingma} \& {Ba}}{{Kingma} \&
  {Ba}}{2014}]{kingma14}
{Kingma} D.~P.,  {Ba} J.,  2014, preprint, \href
  {http://adsabs.harvard.edu/abs/2014arXiv1412.6980K} {} (\mn@eprint {arXiv}
  {1412.6980})

\bibitem[\protect\citeauthoryear{Krizhevsky, Sutskever  \& Hinton}{Krizhevsky
  et~al.}{2012}]{krizhevsky12}
Krizhevsky A.,  Sutskever I.,   Hinton G.~E.,  2012, in Proceedings of the 25th
  International Conference on Neural Information Processing Systems - Volume 1.
  NIPS'12.
Curran Associates Inc., USA, pp 1097--1105, \url
  {http://dl.acm.org/citation.cfm?id=2999134.2999257}

\bibitem[\protect\citeauthoryear{{Larsen}, {Morgan}  \& {Goldstein}}{{Larsen}
  et~al.}{1992}]{larsen92}
{Larsen} J.~T.,  {Morgan} W.~L.,   {Goldstein} W.~H.,  1992, \mn@doi [Review of
  Scientific Instruments] {10.1063/1.1143558}, \href
  {http://ads.nao.ac.jp/abs/1992RScI...63.4775L} {63, 4775}

\bibitem[\protect\citeauthoryear{{Lecun}, {Bengio}  \& {Hinton}}{{Lecun}
  et~al.}{2015}]{lecun15}
{Lecun} Y.,  {Bengio} Y.,   {Hinton} G.,  2015, \mn@doi [\nat]
  {10.1038/nature14539}, \href {http://ads.nao.ac.jp/abs/2015Natur.521..436L}
  {521, 436}

\bibitem[\protect\citeauthoryear{{Mikolov}, {Chen}, {Corrado}  \&
  {Dean}}{{Mikolov} et~al.}{2013}]{mikolov13}
{Mikolov} T.,  {Chen} K.,  {Corrado} G.,   {Dean} J.,  2013, preprint, \href
  {http://adsabs.harvard.edu/abs/2013arXiv1301.3781M} {} (\mn@eprint {arXiv}
  {1301.3781})

\bibitem[\protect\citeauthoryear{Mor{\'e}}{Mor{\'e}}{1978}]{more78}
Mor{\'e} J.~J.,  1978, The Levenberg-Marquardt algorithm: Implementation and
  theory.
Springer Berlin Heidelberg, Berlin, Heidelberg, pp 105--116,
  \mn@doi{10.1007/BFb0067700}, \url {https://doi.org/10.1007/BFb0067700}

\bibitem[\protect\citeauthoryear{Nair \& Hinton}{Nair \& Hinton}{2010}]{nair10}
Nair V.,  Hinton G.~E.,  2010, in Fürnkranz J.,  Joachims T.,  eds,
  Proceedings of the 27th International Conference on Machine Learning
  (ICML-10). Omnipress, pp 807--814, \url
  {http://www.icml2010.org/papers/432.pdf}

\bibitem[\protect\citeauthoryear{{Nandra} et~al.,}{{Nandra}
  et~al.}{2013}]{nandra13}
{Nandra} K.,  et~al., 2013, preprint, \href
  {http://ads.nao.ac.jp/abs/2013arXiv1306.2307N} {} (\mn@eprint {arXiv}
  {1306.2307})

\bibitem[\protect\citeauthoryear{Nelder \& Mead}{Nelder \&
  Mead}{1965}]{nelder65}
Nelder J.~A.,  Mead R.,  1965, Computer Journal, 7, 308

\bibitem[\protect\citeauthoryear{{Shallue} \& {Vanderburg}}{{Shallue} \&
  {Vanderburg}}{2017}]{shallue17}
{Shallue} C.~J.,  {Vanderburg} A.,  2017, preprint, \href
  {http://adsabs.harvard.edu/abs/2017arXiv171205044S} {} (\mn@eprint {arXiv}
  {1712.05044})

\bibitem[\protect\citeauthoryear{{Silver} et~al.,}{{Silver}
  et~al.}{2016}]{silver16}
{Silver} D.,  et~al., 2016, \mn@doi [\nat] {10.1038/nature16961}, \href
  {http://ads.nao.ac.jp/abs/2016Natur.529..484S} {529, 484}

\bibitem[\protect\citeauthoryear{{Smith}, {Brickhouse}, {Liedahl}  \&
  {Raymond}}{{Smith} et~al.}{2001}]{smith01}
{Smith} R.~K.,  {Brickhouse} N.~S.,  {Liedahl} D.~A.,   {Raymond} J.~C.,  2001,
  \mn@doi [\apjl] {10.1086/322992}, \href
  {http://ads.nao.ac.jp/abs/2001ApJ...556L..91S} {556, L91}

\bibitem[\protect\citeauthoryear{{Takahashi} et~al.,}{{Takahashi}
  et~al.}{2016}]{takahashi16}
{Takahashi} T.,  et~al., 2016, in Space Telescopes and Instrumentation 2016:
  Ultraviolet to Gamma Ray. p. 99050U, \mn@doi{10.1117/12.2232379}

\bibitem[\protect\citeauthoryear{{Tanaka} \& {Tomiya}}{{Tanaka} \&
  {Tomiya}}{2017}]{tanaka17}
{Tanaka} A.,  {Tomiya} A.,  2017, \mn@doi [Journal of the Physical Society of
  Japan] {10.7566/JPSJ.86.063001}, \href
  {http://adsabs.harvard.edu/abs/2017JPSJ...86f3001T} {86, 063001}

\bibitem[\protect\citeauthoryear{{Yamada} et~al.,}{{Yamada}
  et~al.}{2016}]{yamada16}
{Yamada} S.,  et~al., 2016, \mn@doi [Journal of Low Temperature Physics]
  {10.1007/s10909-015-1362-2}, \href
  {http://adsabs.harvard.edu/abs/2016JLTP..184..688Y} {184, 688}

\bibitem[\protect\citeauthoryear{{Zeiler} \& {Fergus}}{{Zeiler} \&
  {Fergus}}{2013}]{zeiler13}
{Zeiler} M.~D.,  {Fergus} R.,  2013, preprint, \href
  {http://adsabs.harvard.edu/abs/2013arXiv1311.2901Z} {} (\mn@eprint {arXiv}
  {1311.2901})

\makeatother
\end{thebibliography}








\bsp	
\label{lastpage}
\end{document}